\documentclass[preprint]{aastex}

\begin{document}

\title{Tidally Excited Modes and $\delta$ Scuti Pulsations in the Eclipsing Triple Star IM Persei }
\author{Jae Woo Lee$^{1}$, Kyeongsoo Hong$^{2}$, and Hye-Young Kim$^{3}$}
\affil{$^1$Korea Astronomy and Space Science Institute, Daejeon 34055, Republic of Korea}
\affil{$^2$Institute for Astrophysics, Chungbuk National University, Cheongju 28644, Republic of Korea}
\affil{$^3$Department of Astronomy and Space Science, Chungbuk National University, Cheongju 28644, Republic of Korea}
\email{jwlee@kasi.re.kr}

\begin{abstract}
IM Per is a triple star system whose eclipsing pair masses and radii are accurate to within 1 \%. The TESS light curve of the program 
target exhibits partial eclipses and multiple oscillations with mmag-level amplitudes. It is found that the oscillations affect 
eclipse timing measurements. Binary modeling of the high-quality TESS data indicates that the eclipsing components of the triple system 
are twin dwarfs with parameters of $M_2$/$M_1$ = 0.995, $R_2$/$R_1$ = 0.901, and $\Delta$($T_{\rm eff,1}$--$T_{\rm eff,2}$) = 12 K in 
an eccentric ($e$ = 0.049), detached configuration. The third light of $l_3$ = 0.054 may mostly come from a G-type tertiary companion. 
Our predicted parallax of 1.52 $\pm$ 0.09 mas is concurrent with the Gaia measurement of 1.52 $\pm$ 0.05 mas. 
Multifrequency analysis of the outside-eclipse residuals reveals 22 significant pulsation signals: four in the gravity-mode region 
(0.03$-$2.22 day$^{-1}$) and 18 in the pressure-mode region (9.19$-$25.12 day$^{-1}$). Of the low frequencies, $f_{11}$ and $f_{14}$ 
are orbital harmonics that can be identified as tidally excited modes. The pulsation periods and constants for the high frequencies, and 
the position in the Cepheid instability strip demonstrate that the pulsating component of IM Per is a $\delta$ Sct variable. 
\end{abstract}

\section{INTRODUCTION}

To understand stellar physics in detail, it is essential to robustly determine fundamental stellar parameters. Today we know there 
are at least twice as many binary and multiple stars as there are single stars (Duch\^ene \& Kraus 2013; Murphy et al. 2018), and 
investigations of their features and behavior continue. In eclipsing binaries (EBs), the masses and radii of each component can be 
measured using high-precision light curves and double-lined radial velocities (RVs) to within an uncertainty of less than 1 \%, 
without any assumptions (Torres et al. 2010; Southworth 2015). Such precise measurements allow us to calibrate stellar evolution models. 
At the same time, stellar pulsations can help reveal the interior structure of stars, from their core to surface layers. As a result, 
because EBs with pulsating components provide reciprocal information on both stellar structures and fundamental properties, they are 
excellent targets for asteroseismic modeling. The pulsations in EBs may be influenced by tidal interaction and mass accretion between 
the binary components (Hambleton et al. 2013; Mkrtichian et al. 2018; Bowman et al. 2019).

Data from space photometry such as Kepler (Koch et al. 2010) and TESS (Ricker et al. 2015) is ten to hundred times more precise than 
ground-based photometry. For example, time-series data from space missions exhibit multifrequency oscillations with amplitudes 
down to the micromagnitude ($\mu$mag) level, which were undetectable from the ground (e.g., Lee et al. 2020; Southworth et al. 2020).  
This is the second paper looking for and characterizing pulsation signals in double-lined detached EBs using high-quality TESS photometry 
(Lee \& Hong 2020). Here, we report that IM Per (TIC 117175528; Gaia DR2 445958725070708864; TYC 3323-1123-1; $V\rm_T$ = $+$11.36, 
$(B-V)\rm_T$ = $+$0.78) has a benchmark EB displaying both tidally excited modes and $\delta$ Sct pulsations. For the star system, 
Lacy et al. (2015) obtained $V$-band light curves and double-lined RVs from echelle spectra. These observations and eclipse timings 
were combined to determine the apsidal motion elements and absolute dimensions of the eclipsing pair for the first time. 
The fundamental properties were best-fitted to a 1.2-Gyr isochrone with [Fe/H] = $-$0.22. They reported that the program target was 
composed of a close EB with a period of 2.254 days and a faint third companion barely detected in the spectra. 

This study is organized as follows. The TESS photometry and times of eclipse are reported in Section 2. Sections 3 and 4 detail 
the binary modeling and multifrequency analysis, respectively. The summary and discussion of our results are given in Section 5.

\section{OBSERVATIONS AND ECLIPSE TIMINGS}

The high-precision photometry of IM Per was taken in a 2-min cadence mode during Sector 18 of the TESS mission (Ricker et al. 2015). 
The time-series data were collected by camera 2 from 2019 November 3 to 27 (BJD 2,458,790.66 $-$ 2,458,814.08). We obtained 
the flux measurements labelled \texttt{SAP$_-$FLUX} (simple aperture photometry) and \texttt{PDCSAP$_-$FLUX} 
(pre-search data conditioning) from the MAST Portal\footnote{https://mast.stsci.edu/portal/Mashup/Clients/Mast/Portal.html}. Trends 
in the raw observations were removed by applying a least-squares spline fit to each outside-eclipse light curve of two segments 
separated by a data gap in the midpoint. The normalized fluxes were transformed to magnitudes by utilizing a TESS magnitude of 
$T_{\rm p}$ = $+$10.462 (Stassun et al. 2019). The resultant SAP data is presented in Figure 1 as magnitude versus BJD, 
where short-period oscillations are clearly visible in the outside eclipses. Both the SAP and PDC light curves were examined during 
our data modeling process, and the results were consistent with each other within their error bars. In this article, we present 
the results from the SAP data because of systematic effects somewhat smaller than the PDC data. 

The times of the IM Per eclipses and their errors were derived from individual measurements of the SAP data using the method of 
Kwee \& van Woerden (1956). The TESS timings are given in Table 1, where Min I and Min II denote the primary and secondary minimum 
epochs, respectively. As displayed in Figure 1 and presented in the next sections, our target star is an oscillating EB showing 
multiperiodic pulsations. The pulsation features may cause changes in the eclipse curves, so the mid-eclipse timings could be 
shifted from the real conjunctions. To check this possibility, we recomputed the timings of minimum light from each eclipse after 
subtracting all pulsation modes extracted in Section 4 from the observed data. These are listed in columns (4)$-$(5) of Table 1. 
As shown in the last column of this table, the differences between the minimum times measured from both datasets are all negative 
except for one (BJD 2,458,793.24138). Further, we can see in Figure 2 that these differences are about 11 sec larger on average 
in the secondary eclipses (Min II) than in the primary eclipses (Min I). These results indicate that the eclipse timings calculated 
from the pulsation-subtracted data are later than those from the observed data, and that the timing differences could mainly be caused 
by multiperiodic oscillations in the eclipsing pair of IM Per. 

We obtained the linear ephemeris of the EB system using ten new primary minima in a least-squares solution, as follows:
\begin{equation}
 \mbox{Min I} = \mbox{BJD}~ (2,458,800.004151\pm0.000027) + (2.2542347\pm0.0000080)E,
\end{equation}
where the 1$\sigma$-error values follow each coefficient. The ephemeris is used to calculate the orbital phases of the TESS 
observations. The eclipsing period corresponds to a frequency of $f_{\rm orb}$ = 0.443610 $\pm$ 0.000002 day$^{-1}$.

\section{BINARY MODELING}

The TESS time-series data in Figure 1 displays the partial eclipses of both minima, and the short-term oscillations in 1$-$3 hours 
in the out-of-eclipse parts. The secondary minima are displaced to an orbital phase of around 0.486. The high-quality light curve 
of IM Per was fitted using the Wilson-Devinney (W-D) model (Wilson \& Devinney 1971; van Hamme \& Wilson 2007) in the same scheme 
as that for the pulsating EB CW Cep (Lee \& Hong 2020). For this synthesis, the mass ratio of the EB and the surface temperature of 
the primary star eclipsed at Min I were set to be $q$ = 0.9948 $\pm$ 0.0033 and $T_{\rm eff,1}$ = 7,580 $\pm$ 150 K, respectively, 
from the spectroscopic measurements of Lacy et al. (2015). We adopted a logarithmic limb-darkening (LD) law for the binary components. 
The LD coefficients of $X_{\rm bol}$, $Y_{\rm bol}$, $x_{T_{\rm P}}$, and $y_{T_{\rm P}}$ were initialized from the values of 
van Hamme (1993). The albedos and gravity-darkening exponents were given as $A_1$ = $A_2$ = 1.0 (Rucinski 1969) and 
$g_1$ = $g_2$ = 1.0 (von Zeipel 1924) according to the temperature of each component. The synchronous rotations of $F_1$ = $F_2$ = 1.0 
were used throughout our analysis process. 

The free parameters in our modeling were the reference time ($T_0$) and the eclipsing period ($P$), the inclination angle ($i$), 
the orbital eccentricity ($e$) and the argument of periastron ($\omega$), the secondary's temperature ($T_{\rm eff,2}$), 
the dimensionless potentials ($\Omega_1$, $\Omega_2$), the passband luminosity ($L_1$) and third light ($\ell_3$), and the linear LD 
coefficients ($x_{T_{\rm P_1}}$, $x_{T_{\rm P_2}}$). The initial values for $i$, $e$, $\omega$, $T_{\rm eff,2}$, and $\ell_3$ were 
adopted from Lacy et al. (2015). The light curve modeling was carried out for both datasets: the observed and pulsation-subtracted data. 
The modeling results are shown in Table 2 and the synthetic light curves calculated from the observed data appear as gray lines in 
Figure 1. The residuals between the observations and the eclipse model are displayed in the bottom panel of the figure. 
The uncertainties of the fitted parameters were estimated according to the method used by Southworth et al. (2020) for the TESS data. 
We first calculated various alternative models with different approaches and treatments for the W-D parameters ($q$, $T_{\rm eff}$, $e$, 
$A$, $g$, $F$, $l_3$, $x_{T_{\rm P}}$), and then obtained the parameter errors from the differences between these models and our solutions 
in Table 2 (Lee \& Hong 2020). 

The absolute dimensions of IM Per were computed from our Model 2 parameters for the pulsation-subtracted data and the velocity 
semi-amplitudes ($K_1$ = 123.17 $\pm$ 0.30 km s$^{-1}$ and $K_2$ = 123.80 $\pm$ 0.28 km s$^{-1}$) of Lacy et al. (2015). 
In this calculation, we adopted the bolometric magnitudes of $M_{\rm bol}$$_\odot$ = +4.73 and the effective temperature of 
$T_{\rm eff}$$_\odot$ = 5780 K for Sun. The bolometric corrections (BCs) were used from the $\log T_{\rm eff}$$-$BC relation of 
Torres (2010). The fundamental parameters of each component star are listed in Table 3, along with the measurements of 
Lacy et al. (2015) for comparison. There is a good agreement between these two. 

The distance ($d$) to our program target can be determined using the distance modulus of $V-A_{\rm V}-M_{\rm V}=5\log{d}-5$, where 
$V$ is the apparent magnitude and $A_{\rm V}$ $\simeq$ 3.1$E(B-V)$ is the interstellar extinction. Following the procedure done by 
Lee \& Hong (2020), we obtained the Johnson $V_{\rm system}$ = +11.29 $\pm$ 0.10 and ($B-V$) = +0.66 $\pm$ 0.14 from 
the Tycho-2 magnitudes (H\o g et al. 2000), and the intrinsic ($B-V$)$\rm_0$ = +0.23 $\pm$ 0.03 from the color$-$temperature relation 
(Flower 1996). Then, $E$($B-V$) = +0.43 $\pm$ 0.14. Because $V_{\rm system}$ is the visual magnitude of the triple system including 
the tertiary component, we derived $V$ = +11.33 for the eclipsing pair of IM Per considering the third light ($l_3$ = 0.033) of 
Lacy et al. (2015) in the $V$ band, instead of using ours ($l_3$ = 0.054) from the TESS band. The distance of 660 $\pm$ 40 pc was 
finally determined. This is 94 $\pm$ 61 pc farther than the distance reported by Lacy et al. (2015), but is well matched with 
659 $\pm$ 22 pc taken by the Gaia parallax of 1.517 $\pm$ 0.050 mas (Gaia Collaboration et al. 2018).

\section{PULSATIONAL CHARACTERISTICS}

The eclipsing components of IM Per are located inside the $\delta$ Sct region of the Cepheid instability strip on the HR diagram 
(cf. Lee et al. 2016). Figure 3 shows the residual light curve after removing the binary effects of Model 1 in Table 2 from 
the observed data, which indicates the presence of periodic oscillations with millimagnitude (mmag) level amplitudes. To search for 
the multiperiodic frequencies in our target star, the PERIOD04 software of Lenz \& Breger (2005) was used for the out-of-eclipse part 
(phases 0.066$-$0.414 and 0.558$-$0.934) of the light residuals in the frequency range of 0$-$360 day$^{-1}$. We extracted 
22 significant frequencies based on the iterative prewhitening method (Lee et al. 2014) and the classical criterion of S/N$\ge$4.0 
(Breger et al. 1993). The results are listed in Table 4, where S/N is the signal to noise amplitude ratio for individual frequencies. 
The uncertainties for frequency, amplitude, and phase were obtained according to Kallinger et al. (2008). The Fourier fit on 
the outside-eclipse data was presented as a red curve in Figure 3. The amplitude spectra for IM Per from PERIOD04 are illustrated in 
Figure 4. There were no frequency signals satisfying our criterion at frequencies higher than 30 day$^{-1}$.

The Rayleigh frequency resolution for the TESS observations used is $\Delta f$ = 1/$T$ = 0.043 day$^{-1}$, where the time span of 
the data is $T$ = 22.7 days. We used the frequency separation of $\Delta f$ to find possible multiples and combinations with 
the extracted frequencies. The results of this process are presented in the last column of Table 4. Frequencies $f_4$, $f_{11}$, 
and $f_{14}$ are the orbital frequency ($f_{\rm orb}$) and its multiples, where $f_{11}-5f_{\rm orb}$ = 0.0018 day$^{-1}$ and 
$f_{14}-4f_{\rm orb}$ = 0.0061 day$^{-1}$ are significantly smaller than the Rayleigh criterion $\Delta f$. 
The former frequency can be caused by imperfect light curve modeling and the other two may be the pulsations excited by tidal forces 
in eccentric binary systems (Welsh et al. 2011; Thompson et al. 2012; Hambleton et al. 2013). The lowest frequency $f_9$ appear to be 
a combination term between $f_1$ and $f_2$. From the remaining 18 frequencies and the absolute parameters of the primary component, 
we got the pulsation constants ($Q$) shown in the sixth column of Table 6. The $Q$ values are in a range of 0.014$-$0.037 days and 
correspond to the pressure-mode oscillations of $\delta$ Sct stars with pulsation constants below 0.04 days (Breger 2000).

\section{SUMMARY AND DISCUSSION}

In this article, we aimed to find and characterize the pulsating features present in a best-studied system, using TESS photometric 
observations of IM Per. The following interesting results were obtained.

\begin{enumerate}
\item The time-series data of the program target, observed during Sector 18, exhibit partial eclipses and multiperiodic oscillations 
with mmag-level amplitudes. We applied our binary modeling separately to the observed and pulsation-subtracted data. Agreement 
between them (Model 1 and Model 2) is very good, which means that the light curve parameters are not significantly dependent on these 
pulsations. IM Per is a triple star system with an eccentric detached EB, whose primary and secondary components fill 67.8 \% and 
62.5 \% of their inner critical Roche lobes, respectively. The chief source of $l_3$ = 0.054$\pm$0.008 may be a G-type tertiary component, 
as directly observed in the echelle spectra of Lacy et al. (2015). 

\item The absolute parameters of the inner close pair were obtained by combining our photometric solutions and the spectroscopic orbits 
($K_1$, $K_2$) of Lacy et al. (2015). As a result, the masses and radii of each star were measured with 1.0 \% or better precision. 
The eclipsing components of IM Per are twin dwarf stars with spectral types of about A7 (Pecaut \& Mamajek 2013) and located in 
the $\delta$ Sct instability region (Lee et al. 2016). The EB-based parallax of 1.515 $\pm$ 0.092 mas calculated from our distance 
(660 $\pm$ 40 pc) is in excellent agreement with the Gaia measurements of 1.517 $\pm$ 0.050 mas (Gaia Collaboration et al. 2018).

\item A multifrequency analysis was applied to the out-of-eclipse part of the residual light curve removing the binary effects. 
We detected 22 frequency signals using the iterative prewhitening procedure and the S/N threshold. Among them, $f_4$ and $f_9$ may 
come from incomplete binary modeling and a difference of $f_2-f_1$, respectively. $f_{11}$ and $f_{14}$ are thought to be 
tidal excited pulsations. The other frequencies, their pulsation constants, and the position of the EB components on the HR diagram 
demonstrate that the oscillating component of IM Per is a $\delta$ Sct-type pulsator. 
\end{enumerate}

To find out which component the observed pulsations of IM Per originated from, we examined the residual light curves during the times 
of the primary and secondary eclipses, respectively. Figure 5 show the amplitude spectra for the two cases from the PERIOD04 program. 
In the figure, there are strong alias frequencies, caused by using only in-eclipse data, and spaced by approximately $f_{\rm orb}$. 
The frequencies and amplitudes from the in-eclipses may be affected by the partial eclipse and its relatively short duration in 
the system. We found no conspicuous difference between the amplitude spectra of both eclipses in the pulsation frequencies extracted 
from the outside-eclipse data. At present, it is hard to ascertain which star is pulsating because the binary components of 
IM Per have the same physical properties. Future high-resolution spectra are required to reveal the main source of the multiperiodic 
pulsations from the changes in RVs and line profiles.

\acknowledgments{ }
This paper includes data collected by the TESS mission, which were obtained from MAST. Funding for the TESS mission is provided by 
the NASA Explorer Program. The authors wish to thank the TESS team for its support of this work. We also appreciate the careful reading 
and valuable comments of the anonymous referee. This research has made use of the Simbad database maintained at CDS, Strasbourg, France, 
and was supported by the KASI grant 2020-1-830-08. K.H. was supported by the grants 2017R1A4A1015178 and 2019R1I1A1A01056776 from 
the National Research Foundation (NRF) of Korea.

\clearpage
\begin{figure}
\includegraphics[scale=0.9]{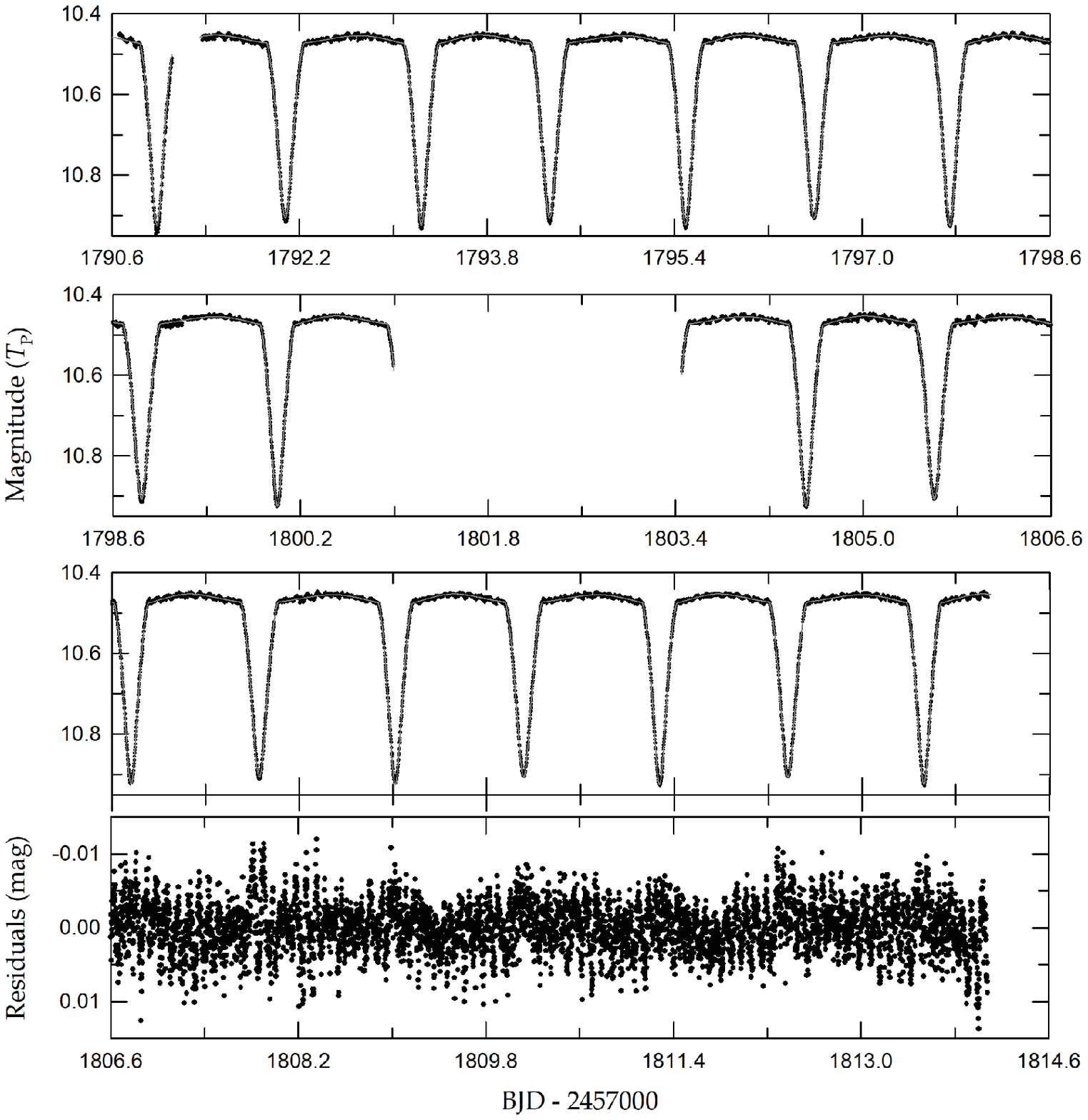}
\caption{The top to third panels show the time-series data of IM Per with the fitted model. The black circles are individual measures 
observed during the TESS Sector 18, and the gray solid lines represent the synthetic curves obtained from our binary star modeling. 
The bottom panel displays a blow-up of the corresponding residuals on the same time baseline as the third panel. }
\label{Fig1}
\end{figure}

\begin{figure}
\includegraphics{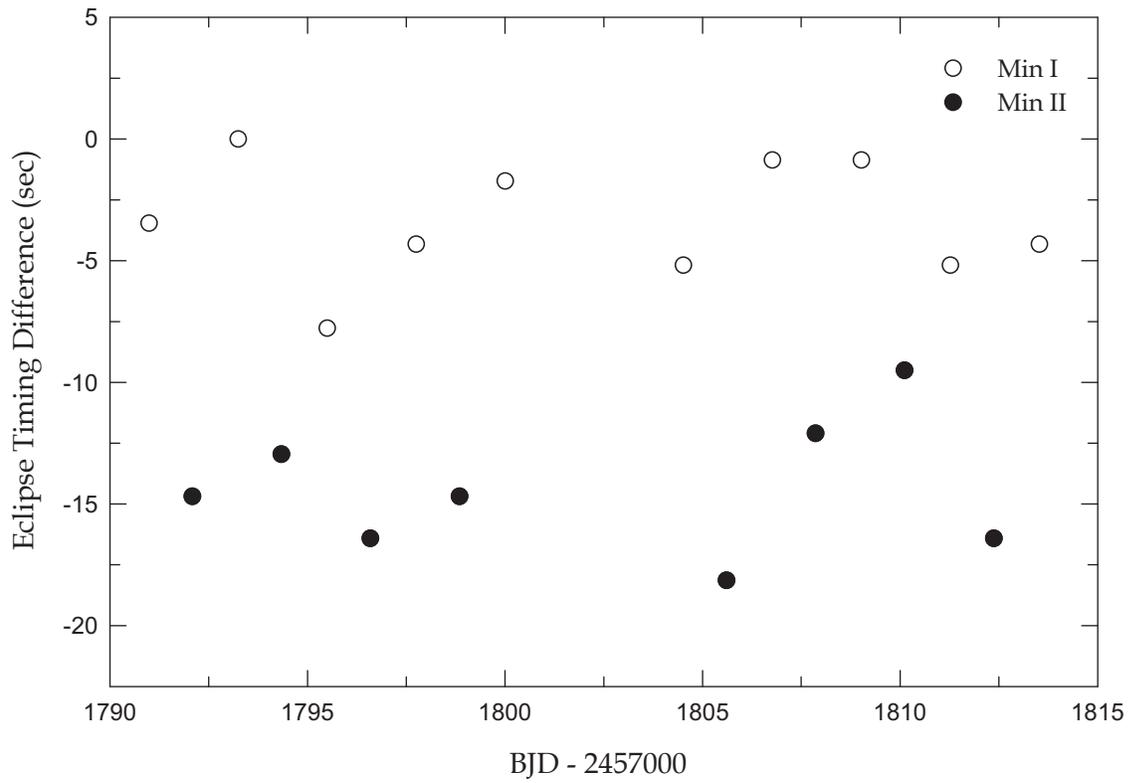}
\caption{Differences between the eclipse times measured from the observed and pulsation-subtracted data. The timing differences show 
the secondary minima (Min II) are later than the primary minima (Min I). }
\label{Fig2}
\end{figure}

\begin{figure}
\includegraphics[scale=0.9]{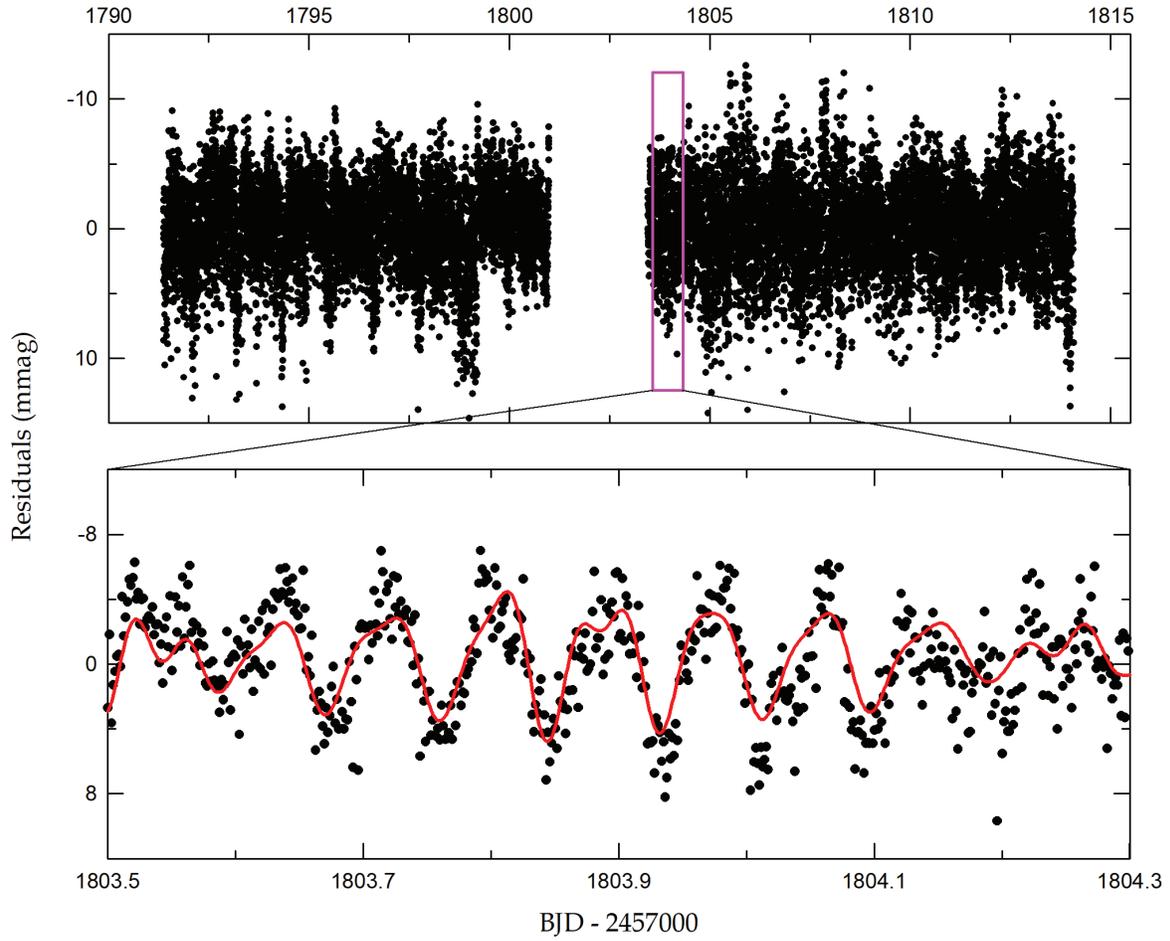}
\caption{Light curve residuals after removing the binarity effects from the observed TESS data. The lower panel presents a short section 
of the residuals which are marked using the inset box in the upper panel. The synthetic curve is computed from the 22-frequency fit to 
the outside-eclipse data. }
\label{Fig3}
\end{figure}

\begin{figure}
\includegraphics[]{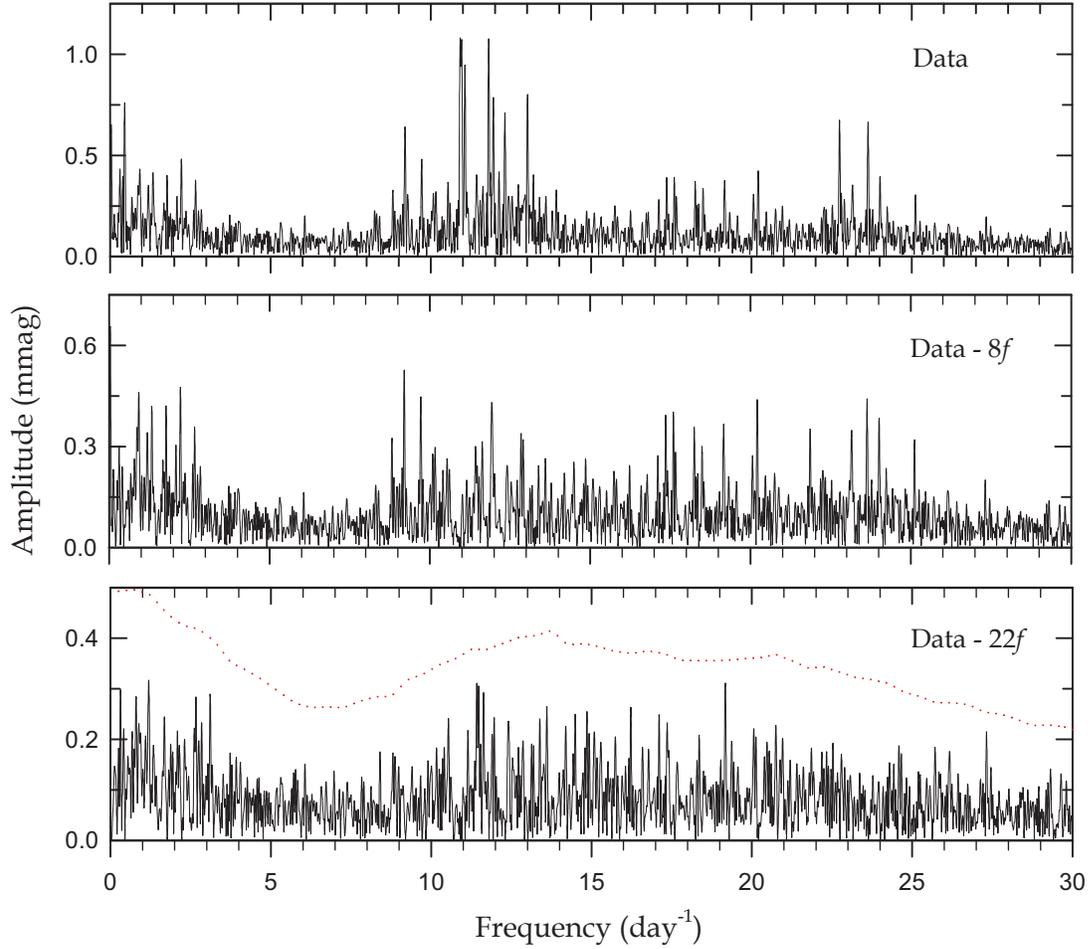}
\caption{Periodogram from the PERIOD04 program for the entire outside-eclipse light residuals. The amplitude spectra before and after 
prewhitening the first eight frequencies and all 22 frequencies are shown in the top to bottom panels. The dotted line in the bottom 
panel corresponds to four times the noise spectrum, which was calculated for each frequency at an equidistant step of 0.5 day$^{-1}$. }
\label{Fig4}
\end{figure}

\begin{figure}
\includegraphics[scale=0.8]{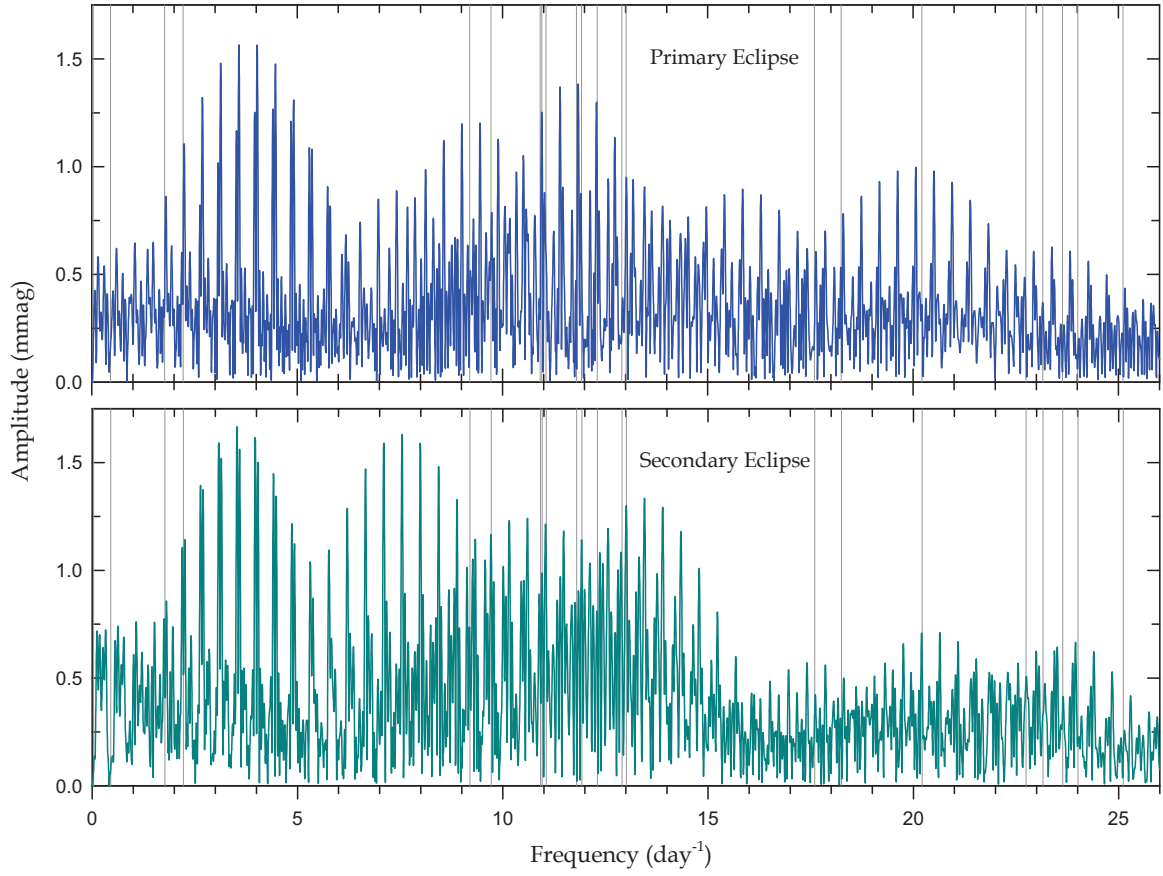}
\caption{Amplitude spectra for the light residuals during times of the primary (upper panel; blue lines) and secondary (lower panel; 
cyan lines) eclipses. In both panels, the vertical gray lines denote the pulsation frequencies found from the outside-eclipse data, 
listed in Table 4. }
\label{Fig5}
\end{figure}

\clearpage 
\begin{deluxetable}{lccccc}
%\tabletypesize{\small} 
\tablewidth{0pt}
\tablecaption{Eclipse Timings of IM Per Measured from Both Datasets }
\tablehead{
\multicolumn{2}{c}{Observed Data}   &                & \multicolumn{2}{c}{Pulsation-subtracted Data}  & \colhead{Difference$\rm ^a$} \\ [1.0mm] \cline{1-2} \cline{4-5} \\[-2.0ex]
\colhead{BJD}    & \colhead{Error}  & \colhead{Min}  & \colhead{BJD}    & \colhead{Error}             &     
}
\startdata                                                                                                         
2,458,790.98754	 & $\pm$0.00012     & I              & 2,458,790.98758	 & $\pm$0.00013               & $-$0.00004         \\
2,458,792.08208	 & $\pm$0.00007     & II             & 2,458,792.08225	 & $\pm$0.00007               & $-$0.00017         \\
2,458,793.24138	 & $\pm$0.00011     & I              & 2,458,793.24138	 & $\pm$0.00009               & ~~0.00000          \\
2,458,794.33614	 & $\pm$0.00006     & II             & 2,458,794.33629	 & $\pm$0.00006               & $-$0.00015         \\
2,458,795.49569	 & $\pm$0.00008     & I              & 2,458,795.49578	 & $\pm$0.00010               & $-$0.00009         \\	     
2,458,796.59026	 & $\pm$0.00006     & II             & 2,458,796.59045	 & $\pm$0.00006               & $-$0.00019         \\
2,458,797.74978	 & $\pm$0.00008     & I              & 2,458,797.74983	 & $\pm$0.00006               & $-$0.00005         \\
2,458,798.84475	 & $\pm$0.00012     & II             & 2,458,798.84492	 & $\pm$0.00011               & $-$0.00017         \\	       
2,458,800.00414	 & $\pm$0.00007     & I              & 2,458,800.00416	 & $\pm$0.00006               & $-$0.00002         \\
2,458,804.51250	 & $\pm$0.00007     & I              & 2,458,804.51256	 & $\pm$0.00008               & $-$0.00006         \\	         
2,458,805.60703	 & $\pm$0.00011     & II             & 2,458,805.60724	 & $\pm$0.00008               & $-$0.00021         \\
2,458,806.76675	 & $\pm$0.00007     & I              & 2,458,806.76676	 & $\pm$0.00008               & $-$0.00001         \\	       
2,458,807.86143	 & $\pm$0.00009     & II             & 2,458,807.86157	 & $\pm$0.00011               & $-$0.00014         \\
2,458,809.02114	 & $\pm$0.00006     & I              & 2,458,809.02115	 & $\pm$0.00006               & $-$0.00001         \\	       
2,458,810.11554	 & $\pm$0.00006     & II             & 2,458,810.11565	 & $\pm$0.00007               & $-$0.00011         \\
2,458,811.27524	 & $\pm$0.00007     & I              & 2,458,811.27530	 & $\pm$0.00007               & $-$0.00006         \\	         
2,458,812.36994	 & $\pm$0.00011     & II             & 2,458,812.37013	 & $\pm$0.00009               & $-$0.00019         \\
2,458,813.52961	 & $\pm$0.00008     & I              & 2,458,813.52966	 & $\pm$0.00008               & $-$0.00005         \\	         
\enddata
\tablenotetext{a}{Differences between columns (1) and (4).}
\end{deluxetable}

\begin{deluxetable}{lccccc}
%\tabletypesize{\small}
\tablewidth{0pt} 
\tablecaption{Binary Parameters of IM Per }
\tablehead{
\colhead{Parameter}                      & \multicolumn{2}{c}{Model 1$\rm ^a$}         && \multicolumn{2}{c}{Model 2$\rm ^b$}         \\ [1.0mm] \cline{2-3} \cline{5-6} \\[-2.0ex]
                                         & \colhead{Primary} & \colhead{Secondary}     && \colhead{Primary} & \colhead{Secondary}         
}
\startdata 
$T_0$ (BJD)                              & \multicolumn{2}{c}{2,458,799.988284(89)}    && \multicolumn{2}{c}{2,458,799.988413(78)}    \\
$P_{\rm orb}$ (day)                      & \multicolumn{2}{c}{2.254226(26)}            && \multicolumn{2}{c}{2.254231(23)}            \\
$q$                                      & \multicolumn{2}{c}{0.9949(33)}              && \multicolumn{2}{c}{0.9949(33)}              \\
$e$                                      & \multicolumn{2}{c}{0.0489(13)}              && \multicolumn{2}{c}{0.0491(10)}              \\
$\omega$ (deg)                           & \multicolumn{2}{c}{117.40(85)}              && \multicolumn{2}{c}{117.13(52)}              \\
$i$ (deg)                                & \multicolumn{2}{c}{83.94(12)}               && \multicolumn{2}{c}{83.927(68)}              \\
$T_{\rm eff}$ (K)                        & 7580(150)         & 7572(150)               && 7580(150)         & 7568(150)               \\
$\Omega$                                 & 5.521(26)         & 5.987(53)               && 5.521(23)         & 5.987(46)               \\
$\Omega_{\rm in}$$\rm ^c$                & \multicolumn{2}{c}{3.742}                   && \multicolumn{2}{c}{3.742}                   \\
$A$                                      & 1.0               & 1.0                     && 1.0               & 1.0                     \\
$g$                                      & 1.0               & 1.0                     && 1.0               & 1.0                     \\
$X_{\rm bol}$, $Y_{\rm bol}$             & 0.671, 0.195      & 0.671, 0.195            && 0.671, 0.195      & 0.671, 0.195            \\
$x_{T_{\rm P}}$, $y_{T_{\rm P}}$         & 0.491(48), 0.210  & 0.457(51), 0.210        && 0.493(28), 0.210  & 0.451(35), 0.210        \\
$l$/($l_{1}$+$l_{2}$+$l_{3}$)            & 0.520(14)         & 0.426                   && 0.5194(97)        & 0.4268                  \\
$l_{3}$$\rm ^d$                          & \multicolumn{2}{c}{0.054(12)}               && \multicolumn{2}{c}{0.0538(75)}              \\
$r$ (pole)                               & 0.2221(14)       & 0.2007(28)               && 0.2221(12)        & 0.2008(21)              \\
$r$ (point)                              & 0.2304(16)       & 0.2061(32)               && 0.2304(14)        & 0.2061(23)              \\
$r$ (side)                               & 0.2246(15)       & 0.2024(29)               && 0.2246(13)        & 0.2024(21)              \\
$r$ (back)                               & 0.2286(16)       & 0.2051(31)               && 0.2286(14)        & 0.2051(23)              \\
$r$ (volume)$\rm ^e$                     & 0.2252(15)       & 0.2028(30)               && 0.2252(13)        & 0.2028(23)              \\ 
%$\sum W(O-C)^2$                          & \multicolumn{2}{c}{0.0029}                  && \multicolumn{2}{c}{0.0025}                  \\ 
\enddata
\tablenotetext{a}{Result from the observed data.}
\tablenotetext{b}{Result from the pulsation-subtracted data.}
\tablenotetext{c}{Potential for the inner critical surface.} 
\tablenotetext{d}{Value at 0.25 phase. }
\tablenotetext{e}{Mean volume radius. }
\end{deluxetable}

\begin{deluxetable}{lcccccccc}
\tablewidth{0pt} 
\tablecaption{Absolute Parameters of IM Per }
\tablehead{
\colhead{Parameter}         & \multicolumn{2}{c}{Lacy et al. (2015)}    && \multicolumn{2}{c}{This Paper}               \\ [1.0mm] \cline{2-3} \cline{5-6} \\[-2.0ex]
                            & \colhead{Primary}  & \colhead{Secondary}  && \colhead{Primary} & \colhead{Secondary}         
}                                                                                             
\startdata
$M$ ($M_\odot$)             & 1.7831$\pm$0.0094  & 1.7741$\pm$0.0097    && 1.7873$\pm$0.0090    & 1.7782$\pm$0.0091     \\
$R$ ($R_\odot$)             & 2.409$\pm$0.018    & 2.366$\pm$0.017      && 2.486$\pm$0.015      & 2.239$\pm$0.025       \\
$\log$ $g$ (cgs)            & 3.9258$\pm$0.0067  & 3.9394$\pm$0.0068    && 3.8990$\pm$0.0057    & 3.9878$\pm$0.0098     \\
$\rho$ ($\rho_\odot$)       &                    &                      && 0.1165$\pm$0.0022    & 0.1587$\pm$0.0055     \\
$T_{\rm eff}$ (K)           & 7580$\pm$150       & 7570$\pm$160         && 7580$\pm$150         & 7568$\pm$150          \\
$\log$ $L$ ($L_\odot$)      & 1.235$\pm$0.035    & 1.217$\pm$0.036      && 1.262$\pm$0.035      & 1.168$\pm$0.036       \\
$M_{\rm bol}$ (mag)         & 1.663$\pm$0.088    & 1.708$\pm$0.091      && 1.575$\pm$0.087      & 1.809$\pm$0.089       \\
BC (mag)                    &                    &                      && 0.033$\pm$0.003      & 0.033$\pm$0.003       \\
$M_{\rm V}$ (mag)           & 1.557$\pm$0.086    & 1.602$\pm$0.090      && 1.542$\pm$0.087      & 1.776$\pm$0.089       \\
Distance (pc)               & \multicolumn{2}{c}{566$\pm$46}            && \multicolumn{2}{c}{660$\pm$40}               \\
\enddata
\end{deluxetable}

\begin{deluxetable}{lrccrcc}
\tablewidth{0pt}
\tablecaption{Results of the multiple frequency analysis for IM Per$\rm ^a$ }
\tablehead{
             & \colhead{Frequency}    & \colhead{Amplitude} & \colhead{Phase} & \colhead{S/N$\rm ^b$}  & \colhead{$Q$}      & \colhead{Remark}    \\
             & \colhead{(day$^{-1}$)} & \colhead{(mmag)}    & \colhead{(rad)} &                        & \colhead{(days)}   &
} 
\startdata 
$f_{1}$      & 10.9142$\pm$0.0019     & 0.73$\pm$0.16       & 4.75$\pm$0.62   &  8.10                  & 0.031              &                     \\
$f_{2}$      & 10.9560$\pm$0.0013     & 1.06$\pm$0.16       & 6.21$\pm$0.43   & 11.65                  & 0.031              &                     \\
$f_{3}$      & 13.0129$\pm$0.0018     & 0.82$\pm$0.17       & 5.67$\pm$0.61   &  8.21                  & 0.026              &                     \\
$f_{4}$      &  0.4515$\pm$0.0030     & 0.62$\pm$0.21       & 0.09$\pm$1.00   &  5.04                  &                    & $f_{\rm orb}$       \\
$f_{5}$      & 12.3082$\pm$0.0020     & 0.74$\pm$0.17       & 5.33$\pm$0.66   &  7.59                  & 0.028              &                     \\
$f_{6}$      & 11.0573$\pm$0.0021     & 0.66$\pm$0.16       & 1.30$\pm$0.70   &  7.16                  & 0.031              &                     \\
$f_{7}$      & 11.8017$\pm$0.0019     & 0.74$\pm$0.16       & 6.27$\pm$0.64   &  7.84                  & 0.029              &                     \\
$f_{8}$      & 22.7467$\pm$0.0024     & 0.52$\pm$0.14       & 4.93$\pm$0.80   &  6.25                  & 0.015              &                     \\
$f_{9}$      &  0.0308$\pm$0.0035     & 0.52$\pm$0.21       & 3.79$\pm$1.17   &  4.28                  &                    & $f_2-f_1$           \\
$f_{10}$     &  9.1987$\pm$0.0022     & 0.55$\pm$0.14       & 3.71$\pm$0.72   &  6.97                  & 0.037              &                     \\
$f_{11}$     &  2.2198$\pm$0.0033     & 0.48$\pm$0.18       & 1.89$\pm$1.10   &  4.55                  &                    & $5f_{\rm orb}$      \\
$f_{12}$     &  9.7162$\pm$0.0027     & 0.47$\pm$0.14       & 6.02$\pm$0.88   &  5.67                  & 0.035              &                     \\
$f_{13}$     & 23.6364$\pm$0.0024     & 0.49$\pm$0.13       & 2.20$\pm$0.81   &  6.18                  & 0.014              &                     \\
$f_{14}$     &  1.7684$\pm$0.0037     & 0.46$\pm$0.19       & 2.85$\pm$1.23   &  4.08                  &                    & $4f_{\rm orb}$      \\
$f_{15}$     & 20.2119$\pm$0.0030     & 0.45$\pm$0.15       & 2.06$\pm$1.01   &  4.95                  & 0.017              &                     \\
$f_{16}$     & 11.9294$\pm$0.0030     & 0.47$\pm$0.16       & 2.87$\pm$1.02   &  4.93                  & 0.029              &                     \\
$f_{17}$     & 17.5979$\pm$0.0033     & 0.41$\pm$0.15       & 1.44$\pm$1.09   &  4.61                  & 0.019              &                     \\
$f_{18}$     & 18.2476$\pm$0.0035     & 0.39$\pm$0.15       & 1.59$\pm$1.15   &  4.35                  & 0.019              &                     \\
$f_{19}$     & 24.0108$\pm$0.0032     & 0.37$\pm$0.14       & 4.07$\pm$1.08   &  4.64                  & 0.014              &                     \\
$f_{20}$     & 12.9072$\pm$0.0044     & 0.34$\pm$0.17       & 3.74$\pm$1.47   &  3.40                  & 0.026              &                     \\
$f_{21}$     & 23.1585$\pm$0.0035     & 0.34$\pm$0.14       & 0.69$\pm$1.17   &  4.29                  & 0.015              &                     \\
$f_{22}$     & 25.1141$\pm$0.0033     & 0.33$\pm$0.12       & 2.66$\pm$1.09   &  4.61                  & 0.014              &                     \\
\enddata                                                                                                                           
\tablenotetext{a}{Frequencies are listed in order of detection. }
\tablenotetext{b}{Calculated in a range of 5 day$^{-1}$ around each frequency. }
\end{deluxetable}

\end{document}